\documentclass{article}

\usepackage[final, nonatbib]{nips_2023}

\usepackage[utf8]{inputenc} 
\usepackage[T1]{fontenc}    
\usepackage{hyperref}       
\usepackage{url}            
\usepackage{booktabs}       
\usepackage{amsfonts, amsmath}       
\usepackage{nicefrac}       
\usepackage{microtype}      
\usepackage{graphicx}
\usepackage{bm}
\usepackage{multirow}
\usepackage{tabularx}
\newtheorem{theorem}{Theorem}[section]

\usepackage{relsize}
\usepackage{makecell}

\title{Non-negative isomorphic neural networks for photonic neuromorphic accelerators}

\author{%
Manos Kirtas$^{1*}$ \quad Nikolaos Passalis$^{1}$ \quad Nikolaos Pleros$^{2}$  \quad Anastasios Tefas$^{1}$\\
$^1$Computational Intelligence and Deep Learning Group\\
\quad $^2$Wireless and Photonic Systems and Networks Group\\
\textit{Dept. of Informatics}, \textit{Aristotle University of Thessaloniki},\\ Thessaloniki, Greece \\
\{eakirtas, passalis, npleros, tefas\}@csd.auth.gr
}
\begin{document}

\maketitle

\begin{abstract}
Neuromorphic photonic accelerators are becoming increasingly popular, since they can significantly improve computation speed and energy efficiency, leading to femtojoule per MAC efficiency. However, deploying existing DL models on such platforms is not trivial, since a great range of photonic neural network architectures relies on incoherent setups and power addition operational schemes that cannot natively represent negative quantities. This results in additional hardware complexity that increases cost and reduces energy efficiency. To overcome this, we can train non-negative neural networks and potentially exploit the full range of incoherent neuromorphic photonic capabilities. However, existing approaches cannot achieve the same level of accuracy as their regular counterparts, due to training difficulties, as also recent evidence suggests. To this end, we introduce a methodology to obtain the non-negative isomorphic equivalents of regular neural networks that meet requirements of neuromorphic hardware, overcoming the aforementioned limitations. Furthermore, we also introduce a sign-preserving optimization approach that enables training of such isomorphic networks in a non-negative manner.
\end{abstract}

\section{Introduction}
\label{sec:intro}

Neuromorphic architectures have gained increasing attention recently, as they provide novel electronic solutions focusing on memory architectures suitable for high-speed matrix-based calculations, which cover a significant fraction of the calculations involved during the inference of Deep Learning (DL) models, with low energy consumption~\cite{pei2019towards, li2020power}. Neuromorphic photonics are among the most promising approaches, with recent layouts already paving a realistic road map towards femtojoule per MAC efficiencies~\cite{9006831}, leveraging advantages in materials and waveguide technologies~\cite{cheng2017chip, feldmann2019all}, enabling ultra-fast analog processing and vector-matrix multiplication with almost zero power consumption~\cite{shen2017deep, 9605987}, significantly exceeding their electronic counterparts~\cite{Murmann2021}.

However, integrating DL models into physically implemented devices comes with additional cost if their physical properties are not taken into account during the implementation phase. For example, the vast majority of currently available photonic architectures relies on incoherent layouts and is facing challenges to support negative quantities, since optical signals get naturally converted into power signals during the nonlinear process that has to take place at the activation stage, implying that the sign information of the weighted sum is ignored. This mechanism turns the use of negative number representations within a Photonic Neural Network (PNN) into a challenging process, typically enforcing the adoption of higher complexity hardware architectures, such as balanced photodetector schemes~\cite{shastri2021photonics}, biasing configurations~\cite{mourgias_iq} and signal transformation blocks~\cite{Mourgias-Alexandris:22}.
%

Our main contributions are two-fold:
\begin{itemize}
    \item We propose a method to transform trained Artificial Neural Networks (ANNs) to their fully non-negative equivalent.
    \item We propose an optimization method which ensures that the model's parameters will remain non-negative during training, enabling one to either train a non-negative model from scratch or continue training in a non-negative manner. 
\end{itemize}

\section{Related Works}
\label{sec:related_works}

\paragraph{Neuromorphic Photonics} PNNs deployment requires the employment of optically enabled mechanisms and photonic building blocks to realize the respective signals and parameters of the neural layer. Input signals are typically imprinted in the optical domain using optical modulators, while weighting functionality generally requires the use of variable optical attenuation schemes~\cite{tait2017neuromorphic}. Several approaches have been utilized to date in on-chip weight implementations of photonic neural layers, including tunable i) optical filtering mechanisms~\cite{tait2017neuromorphic, Mourgias-Alexandris:22, 9489835}, ii) waveguide absorption techniques~\cite{cheng2017chip, feldmann2019all} and, iii) optical gain approaches~\cite{10.1117/12.2587668, 9748238}. Summation is then performed in the optical domain via i) wavelength multiplexers or power combiners in the case of incoherent layouts~\cite{feldmann2019all, 9557801, shastri2021photonics} and ii) interferometric stages and optical couplers in the case of coherent architectures~\cite{shen2017deep, 2023963973, 9605987, 9748238}. Finally, the non-linear activation can be offered by either i) optoelectronic schemes~\cite{feldmann2019all, 2023963973, 9489835} or ii) all-optical non-linear modules~\cite{feldmann2019all, 2023963973, 9489835}.

 Furthermore, there are several works that take into account the unique nature of neuromorphic photonics and design the training and deployment of models accordingly~\cite{mourgias2022noise, 9606097, Mourgias-Alexandris:22}. Although such approaches integrate transfer function- and noise-related limitations of photonic hardware~\cite{9815870, 9287649, 9489835}, leading to significant performance improvements during deployment, they typically ignore the sign limitation of photonic architectures.

\paragraph{Cost Reduction} Decreasing the hardware complexity of PNNs has been extensively studied in the literature, and several approaches have been proposed, ranging from pruning~\cite{9557801}, to quantization~\cite{KIRTAS2022561}, and mixed-precision representations~\cite{2023963973}. However, even the simplest PNN implementation would require amplitude modulators for its inputs, meaning that in this case both the input signal modulators and photonic weighting stage control just the amplitude of the optical field, resulting in non-negative networks~\cite{9006831}. Introducing sign information in a PNN has to incorporate a new physical dimension that can be used to correlate its state with the sign. Coherent photonic architectures offer a rather simple way of representing signed optical signals by correlating the sign to the phase of the propagating optical fields~\cite{9605987, Mourgias-Alexandris:22, 9489835}. However, this requires additional phase modulation circuitry both at the input signal generation and weighting stage, also requiring more complex circuitry at the non-linear activation stage in order to account for the sign information of the weighted optical sum at the activation unit, such as an optical biasing scheme or a coherent receiver~\cite{9606046, 9748238, 9605987, 9557801}. 

\paragraph{Non-negative ANNs} Although there are some works studying non-negativity on ANNs, they mostly target partial nonnegative architectures that are focused on reconstruction (e.g. autoencoders)~\cite{8051252, Chorowski2015a} applied on small datasets or non-traditionally used ANNs (such as Pyramid Neural Networks)~\cite{8489216}, facing difficulties in scaling ability, generalization in DL architectures (such as Convolutional Neural Networks (CNNs), Recurrent Neural Networks (RNNs)) and result in significantly performance degradation that hinders their application on both DL and neuromorphic photonics. Our proposed framework can be applied on traditionally used architectures without any major change and performance degradation, since it is based on the non-negative isomorphic representation of traditional models.   

\paragraph{Non-negative Training} Existing training methods, which are oriented to non-negative architectures, are based on either limited memory optimization (such as Quasi-Newton)~\cite{8051252, Chorowski2015a} or in non-gradient based optimization methods~\cite{8489216}, clipping parameters during the backpropagation, constraining in such a way the variance of parameters during the first epochs of training, which leads on convergence difficulties (typical examples are demonstrated in the Appendix). Our proposed non-negative optimization method is based on a multiplicative alternative of the Stochastic Gradient Descent (SGD) optimizer that combined with the proposed non-negative transformation ensures that the variance of parameters will not diminish, allowing the training process to proceed smoothly without reducing the variance. 
 
Although multiplicative updates have been extensively studied during the early years of machine learning research~\cite{v008a006, KIVINEN1997325}, to the best of our knowledge, this is the first work that investigates them in the context of non-negative training. Even works that target sign-preserving optimization are limited to studying both the excitatory and inhibitory functions of neurons, obeying in general Dale's rule~\cite{eccles2013physiology}, pointing out mostly the anatomical correlation with biological synapses~\cite{Amit_1989, 10.7554}. In such a direction is also the recent work~\cite{9a32ef65} that leverages multiplicative updates on Adam to train lower bit-width synapses stored in logarithmic numbers oriented to software-hardware co-design. The authors note that the sign-pattern of initialized weights can possibly restrict the expressive ability of networks. Our work goes beyond such approaches since we claim that even with positive sign only parameters, we can acquire an expressive isomorphic representation of a network by applying the appropriate transformation and training it in a non-negative manner. 

\paragraph{Isomorphism in ANNs} Isomorphism is a general mathematical property that is especially useful in graph theory. As a result, there are several works that consider isomorphism to extract more expressive representations of the input graphs on graph classification problems~\cite{9721082, meng2019isonn}. As far as we know, this is the first work studying the isomorphism of the networks and proposes a structured methodology to acquire their non-negative equivalent opening a new research direction with possible wider implications, e.g., design of isomorphic networks that are adjusted for conventional accelerators as well, providing potentially more explainable DL architectures~\cite{8051252, Chorowski2015a, 8489216}.

\section{On the Difficulty of Training Non-negative Neural Networks}
\label{sec:on_difficulty}

\begin{figure*}
    \centering
    \includegraphics[width=\linewidth]{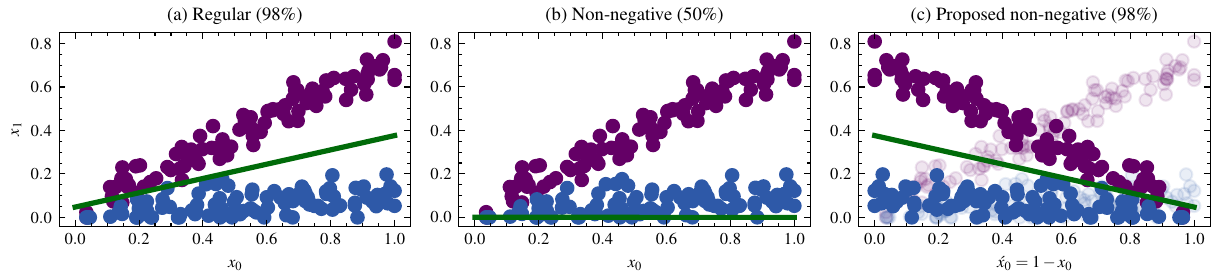}
    \caption{(a) The resulting positive-slope decision boundary as acquired by the traditional optimization process. (b) Regular training of the non-negative classifier results in a non-positive slope decision boundary. (c) The proposed transformation on the trained parameters of (a).}
    \label{fig:tranformation}
\end{figure*}

Conceptually, ANNs project the input features to a latent dimensional space, aiming to represent them in a more separable way, seeking for a hyperplane decision boundary that can classify them in an optimal way. However, classifying non-negative features even when they are linear separable is often impossible when using non-negative parameters to a linear classifier. For example, in a two dimensional binary classification task, as depicted in Figure~\ref{fig:tranformation}.a, the decision boundary of a traditional logistic regression classifier is given by:
\begin{equation}
    x_1 = - \frac{w_0}{w_1}x_0 - \frac{b}{w_1} \in \mathbb{R},
    \label{eq:reg_db}
\end{equation}
where \(x_i \in [0, 1]\) is the input, \(w_i \in \mathbb{R}\)  and \(b\in \mathbb{R}\) are the weights and biases of the classifier respectively and \(i \in \{0, 1\}\). We can easily conclude that there is a positive slope decision, with the slope given by \(m=-w_{0}/w_{1}\), which requires a negative weight. In fact, training the linear classifier, using SGD, we obtain a positive slope decision boundary, as shown in Figure~\ref{fig:tranformation}.a, achieving optimal performance with \(w_0<0\). On the other hand, constraining the classifier to non-negative parameters results in a negative slope decision boundary that is unable to discriminate the two classes, as presented in Figure~\ref{fig:tranformation}.b.

Inspired by the fact that challenging computational tasks, such as calculating the motion of our solar system's planets in a geocentric manner, can be easily solved by changing the coordinate system, for example, calculating the motion of the planets on a heliocentric system, we considered isomorphism, i.e. the same behavior but with different implementation and/or parameters, to claim that an equivalent classifier with non-negative parameters exists. Such a non-negative isomorphic classifier produces the same classification outcome as the original one by applying a coordination change to the original inputs and parameters. Indeed, in the example of Figure~\ref{fig:tranformation}.a, we can easily transform the original problem and classifier, by shifting and rotating points and decision boundary, getting an equivalent non-negative classifier, as depicted in Figure~\ref{fig:tranformation}.c, using only positive parameters. Such non-negative isomorphic classifier outcomes the decision boundary given by:
\begin{equation}
    x_1 =-\frac{\left|w_{0}\right|}{w_{1}} x_0' -\frac{b-a\left|w_{0}\right|}{w_{1}} \in \mathbb{R}
\end{equation}
where \(w_0 < 0\), \(w_1 > 0\), \(a=1\) and \(x_0'=\left(a-x_0\right)\). The non-negative classifier leads to a rotated decision boundary and is equal to the original classifier performance.



\section{Non-negative Isomorphic Neural Networks}
\label{sec:tranformation}
\paragraph{Definition 4.1 (Linear Neuron)}Let \(z_i \in \mathbb{R}\) the response of the linear part of the \(i\)-th neuron of a fully connected layer, where \(i = 1 \ldots N\), given by: 

\begin{equation}
\label{eq:linear_pass}
  z_i = u_i(\bm{x}) = \sum_{j=1}^{M}{w_{ij}} x_j + b_i  \in \mathbb{R}, 
\end{equation}
where \(\bm{w}_i \in \mathbb{R}^{M}\), \(b_i \in \mathbb{R}\) and \(\bm{x} \in \mathbb{R}^{M}\) are the weight, bias and input vectors of the \(i\)-th neuron, respectively. Assuming an activation function \(g(\cdot): \mathbb{R} \rightarrow \mathbb{R}_{+} \), where \(\mathbb{R}_{+}\) denotes the set of positive real values, the outputs of \(i\)-th neuron is given by:
\begin{equation}
    y_i = g(z_i) \in \mathbb{R}_{+}.
\end{equation}

\begin{theorem}
\label{theo:transformation}
For every linear neuron given by Definition 4.1 there is a non-negative isomorphic with the linear response provided as:
\begin{equation}
     u_i'(\bm{x}') = z_i' =\sum_{j=1}^{M}|w_{ij}|x'_j + b'_i \in \mathbb{R}_+,
\end{equation} and the output as: 
\begin{equation}
   y_i' = g_c(u_i'(\bm{x}')) \in \mathbb{R}_+,
\end{equation} where:
\begin{equation}
    x'_j= 
    \begin{cases}
        a - x_j & \text{if } w_{ij} <  0\\
        x_j     & \text{otherwise}
    \end{cases}.
\end{equation}
Then, appropriate parameters \( b_i' \in \mathbb{R}_+ \) and \( a \in \mathbb{R}_+ \), as well as activation function \( g_c(\cdot): \mathbb{R} \rightarrow \mathbb{R}_+\) exist so the neuron leads to the same response, i.e., \(y_i = y'_i\).
\end{theorem}


\paragraph{Proof 4.1:} Equation~\ref{eq:linear_pass} can be written as: 
\begin{equation}
\label{eq:transormation_1}
  z_i = -\sum_{\{j | w_{ij} < 0\}}|w_{ij}| x_j + \sum_{\{j | w_{ij} > 0\}}  w_{ij} x_{j} + b_i \in \mathbb{R}.
\end{equation}
Assuming that every input of a linear layer is non-negative, which can be enforced in the input of the network by trivially performing normalization to features, then, by adding and subtracting the quantity \(\alpha\sum_{\{j | w_{ij} < 0\}}{|w_{ij}|}\), where  \(a \geq max(\mathcal{\bm{X}})\) and \textit{max} denotes the maximum element in the feasible set \(\{\bm{x} : \bm{x} \in \mathbb{R}_{+}^{N} \} \), defined as \(\mathcal{X}\), the Equation~\ref{eq:transormation_1} can be written as:
\begin{equation}
\label{eq:transformation_2}
\begin{gathered}
   z'_i = \alpha\sum_{\{j | w_{ij} < 0\}}{|w_{ij}|} - \sum_{\{j | w_{ij} < 0\}}{|w_{ij}|x_j} + \sum_{\{j |w_{ij} > 0\}}{w_{ij}x_{j}} + \left(b_i - \alpha\sum_{\{j |w_{ij} < 0\}}|w_{ij}|\right) \in \mathbb{R}.
\end{gathered}
\end{equation}
This allows us to rotate the input feature space, similarly to Section~\ref{sec:on_difficulty}. To this end, the first two terms can be merged, while the last term can be integrated to an updated bias term:
\begin{equation}
\tilde{b}_i = b_i - \alpha\sum_{\{j |w_{ij} < 0\}}|{w}_{ij}| \in \mathbb{R}.
\end{equation}
Therefore, Equation~\ref{eq:transformation_2} can be written as:
\begin{equation}
\label{eq:transformation_3}
     z'_i = \sum_{\{j |w_{ij} < 0\}}|w_{ij}|(a - x_{j}) + \sum_{\{j |w_{ij} > 0\}}w_{ij}x_j + \tilde{b}_i \in \mathbb{R}.
\end{equation}
To simplify the Equation~\ref{eq:transformation_3}, we can define the rotated input as:
\begin{equation}
    x_j' = f(x_j) = 
    \begin{cases}
        a - x_j & \text{if } w_{ij} <  0\\
        x_j     & \text{otherwise},
    \end{cases}
\end{equation}
where \(x_j' \in \mathbb{R}_{+}\). Similarly, the updated non-negative weights of the \(i\)-th neuron can be directly calculated as:
\begin{equation}
    \bm{w}'_{i} = |\bm{w}_i| \in \mathbb{R}_{+}^{M},
\end{equation}
where \(|\cdot|\) denotes the element-wise absolute value, i.e., \(w'_{ij} = \{|w_{ij}|, j=0\ldots M\}\), since all weights involved in (\ref{eq:transformation_3}) are positive. Furthermore, the new non-negative biases are transformed according to the following formula:
\begin{equation}
    b^{'}_i = \tilde{b}_{i} + c \in \mathbb{R}_{+},
\end{equation}
where  \(c\) is computed as:
\begin{equation}
    c = max\{|\tilde{b}_0|, \ldots |\tilde{b}_{N}|\} \in \mathbb{R}_{+},
\end{equation}
denoting \textit{activation shifting point}. The \textit{activation shifting point} is applied to the original activation to slide it into the input domain, leading to the same output as the original network. To ensure that the network will work in a non-negative manner, the original activation function has to work on a non-negative output space, i.e., \(g(z): \mathbb{R}\rightarrow [g_{min}, g_{max}]\), where \( 0 \leq g_{min} < g_{max} < \infty\), with the shifted activation is calculated as:
\begin{equation}
    g_c(x)  = g(x - c) \in \mathbb{R}_{+}.
    \label{eq:nn_activation}
\end{equation}
Typically the \(a^{(k+1)}\) of the next layer, where the \(k\) is the number of current layer, can be set to \(g_{max}\), validating the aforementioned assumption, \(a^{(k+1)} = g_{max} = max(\mathcal{\bm{X}})\). 

Since Non-negative Transformation (NNT) targets neuromorphic architectures, the \textit{activation shifting} can be integrated during the design phase of the activation function, allowing one to integrate it in layer. This evaluates Theorem~\ref{theo:transformation} since the original parameters are transformed to their non-negative equivalent, with the isomorphic neuron has the same response as the original one.


The computational complexity of the proposed transformation is equal to the number of trainable parameters and can be trivially acquired by taking into account its algorithmic representation, presented in Appendix. More precisely, if the \(\bm{\theta}^{(k)}\) defines the trainable parameter of \(k\)-th layer, consisting of weights \(\bm{w}^{(k)}\) and biases \(\bm{b}^{(k)}\), with \(k=1\ldots n\), then the computational complexity of the transformation is \(O(|\bigcup\{\bm{\theta}^{(k)}\}_1^{n}|)\), where \(|\cdot|\) denotes the number of elements of the corresponding set.


\paragraph{Definition 4.2 (Recurrent Neuron)} Let \(\bm{x}_t \in \mathbb{R}^M\) denote \(M\) features fed on \(i\)-th neuron at \(t\)-th time-step. Each recurrent neuron processes two signals: a) the current input signal, which is weighted by \(\bm{w}_i^{(in)} \in R^{M}\), and b) a recurrent feedback signal, denoted by \(\mathbf{y}_{t-1}^{(r)} \in R^N\) and weighted by a set of recurrent weights \(\bm{w}_i^{(r)} \in R^{N}\), which corresponds to the output of the \(N\) recurrent neurons of the same layer in a previous time step. The linear response of \(i\)-th neuron is given by:
\begin{equation}
\label{eq:rnn}
      u_{ti}^{(r)}(\bm{x}_t, \bm{y}_{t-1}) = \sum_{j=1}^{M} w_{ij}^{(in)}x_{tj} +  \sum_{j=1}^{N} w_{ij}^{(r)}y^{(r)}_{t-1,j} + b_{i} \in \mathbb{R}_+,
\end{equation} and the outputs as:
\begin{equation}
   y_{ti}^{(r)} = g(u_{ti}^{(r)}(\bm{x}_t, \bm{y}^{(r)}_{t-1})) \in \mathbb{R}_+, 
\end{equation}

\begin{theorem}
\label{theo:transformation_rnn}
For every recurrent neuron given by Definition 4.2 there is a non-negative isomorphic with the linear response provided as:
\begin{equation}
       z_{ti}^{(r)'} = u_{ti}^{(r)'}(\bm{x}'_t, \bm{y}'_{t-1}) = \sum_{j=1}^{M} |w_{ij}^{(in)}|x'_{tj} +  \sum_{j=1}^{N} |w_{ij}^{(r)}|y^{(r)'}_{t-1,j} + b'_{i} \in \mathbb{R}_+,
\end{equation} and the output as: 
\begin{equation}
   y_{ti}^{(r)'} = g_c(u_{ti}^{(r)'}(\bm{x}'_t, \bm{y}^{(r)'}_{t-1})) \in \mathbb{R}_+, 
\end{equation}where:
\begin{equation}
    x'_{tj}= 
    \begin{cases}
        a^{(in)} - x_{tj} & \text{if } w_{ij}^{(in)} <  0\\
        x_{tj}    & \text{otherwise}
    \end{cases},
\end{equation} and:
\begin{equation}
    y^{(r)'}_{t-1,j}= 
    \begin{cases}
        a^{(r)} - y^{(r)}_{t-1,j} & \text{if } w_{ij}^{(r)} <  0\\
        y^{(r)}_{t-1,j}    & \text{otherwise}
    \end{cases}.
\end{equation} 
Then, appropriate parameters \(b'_i \in \mathbb{R}_+\), \( a^{(in)} \in \mathbb{R}_+ \) and \( a^{(r)} \in \mathbb{R}_+ \), as well as activation function \( g_c(\cdot): \mathbb{R} \rightarrow \mathbb{R}_+\) exist so the neuron leads to the same response, i.e., \(y_{ti}^{(r)} = y_{ti}^{(r)'} \).
\end{theorem}

\paragraph{Proof 4.2:} Recurrent neurons can conceptually be seen as two linear neurons in which the current input signal, \(x_{tj}\), is fed to one neuron and the recurrent feedback signal, \(y^{(r)}_{t-1,j}\), to the other. To this end, we can easily conclude that there is a non-negative isomorphic for recurrent neurons by applying Proof 4.1 on the first two terms of Equation~\ref{eq:rnn}.


\paragraph{Definition 4.3 (2D Convolutional Neuron)} Let \(\bm{X} \in {\mathbb{R}}^{C \times N \times M}\) denote the input of a 2D convolutional layer, where \(C\), \(N\), and \(M\) represent the number of channels, height and width of the input feature, with the layer consist of the kernel's weights \(\bm{W} \in {\mathbb{R}}^{D \times C \times N^k \times M^k}\), where \(D\) denotes the number of output channels, while \(N^k\) and \(M^k\) represent the height and width of the kernel. The bias is denoted as \(\bm{b} \in {\mathbb{R}}^{D}\). Convolutional neurons can be constructed by trivially extending Definition 4.1 by sliding the linear neuron over the input after flattening each input patch to a vector. Thus, starting from the \(c\)-th channel of the input and flattening the \(i\)-th \(N^k \times M^k\) sub-matrix of \(\bm{X} \in {\mathbb{R}}^{C \times N \times M}\), denoted as \(\bm{x}_{i} \in \mathbb{R}^{N^k M^k}\), we can define the \(z_{i}\) the linear output of the \(i\)-th element of \(d\)-th output channel given by: 
\begin{equation}
\label{eq:conv_pass}
  z_{i}= u_i(\bm{x}) = \sum_{j=1}^{M}{w_{j}} x_{ij} + b_d  \in \mathbb{R}, 
\end{equation}
where \(\bm{w} \in \mathbb{R}^{N^k M^k}\) is the equivalent to the flattened kernel weight of \(\bm{W}_{cd} \in \mathbb{R}^{N^k_i \times M^k_i}\), where \(c\) and \(d\) are the current input and output channels, respectively. To this extent, Theorem 4.1 can also be applied to convolutionals, since they can be defined as linear neuron building blocks, allowing us to apply Proof 4.1. Without loss of generality the theorem can be applied to 1D, 3D and multidimensional convolutionals as well.
\section{Non-negative Optimization}
\label{sec:nn_optimization}

Transforming an already trained network to its non-negative isomorphic can be limiting in cases where continuing training is required. To this end, we propose an adjustments on SGD that constrains the trainable parameter \(\bm{\theta}^{(k)}\) of \(k\)-th layer, consisting of weights \(\bm{w}^{(k)}\) and biases \(\bm{b}^{(k)}\), to non-negative quantities. More precisely, we modify the additive update rule, in which sign shifting is attributed during SGD optimization, by normalizing the gradients using the non-linear function \(\tanh: \mathbb{R} \rightarrow (-1, 1)\) and multiplying it by the absolute value of the parameter. The optimization algorithm picks a point \( \bm{\theta}_\tau^{{(k)}}\), at each time step \(\tau\), and updating it according to:
\begin{equation}
\bm{\theta}_{\tau}^{(k)} = \bm{\theta}_{\tau-1}^{(k)} +
|\bm{\theta}^{(k)}_{\tau-1}|\tanh{\left(-\eta_{in}\frac{\partial J}{\partial \bm{\theta}^{(k)}_{\tau-1}} \right)} \eta_{out} \in \mathbb{R}^{+},
\label{eq:m_abs}
\end{equation}
where \(\eta_{in} \in \mathbb{R}^{+}\) is the inner and \(0 < \eta_{out} \leq 1\) the outer learning rate. Essentially, the inner learning rate allows one to adjust the gradients regarding the working range of the used non-linearity. The outer learning rate affects the size of the step similarly to the learning rate used in traditionally applied optimization methods. The proposed optimization method is a sign-preserving alternative of SGD, named \textit{non-negative stochastic gradient descent} (NNSGD), and when combined with the NNT can be used to post-train or train from scratch DL models in a non-negative manner. Both proposed non-negative optimization is presented algorithmically as well in the Appendix. 
\section{Experimental Evaluation}
\label{sec:exprimental_results}

We experimentally evaluate the proposed framework using a wide range of architectures and photonic configurations demonstrating its capabilities in three scenarios: a) transforming pretrained ANN to its non-negative isomorphic without any performance degradation, b) continuing training of a non-negative isomorphic network that is regularly pretrained, and c)  non-negative training from scratch. We evaluate it in image classification (MNIST, Fashion MNIST, CIFAR10), malware classification (Malimg~\cite{malimg}), a large scale financial time-series forecasting (FI2010~\cite{8713851}) and a simple natural language processing (Names~\cite{names_dataset}) task. Details about the experimental setup, models applied, datasets, photonic configurations and hyper parameter tuning process are provided in detail in the Appendix. Note that we applied the proposed framework to small models according to current neuromorphic photonics capabilities and limitations. Additionally, the employed hyper parameter tuning is applied to all scenarios for fairness since both proposed methods and evaluated baselines are targeting non-negative quantities with significantly different parameter distributions and magnitudes with those the SGD traditionally targets.    

\subsection{Trained model transformation}

\begin{table}[]
    \caption{Applying proposed transformation in various datasets, architectures and photonic configurations. The Non-negative Match column defines if the performance of the non-negative isomorphic matches trained model's one. \label{tab:res_trans}}
    \begin{center}
    \resizebox{\linewidth}{!}{
    \begin{tabular}{l|l|lc||lc}
    \multirow{2}{*}{Dataset} & \multirow{2}{*}{Architecture} & \multicolumn{2}{c||}{Photonic Sigmoid} & \multicolumn{2}{c}{Photonic Sinusoidal} \\
    \cline{3-6}
     & & Regular & Non-negative Match & Regular & Non-negative Match \\
    \midrule
    MNIST & MLP & \(97.44\pm{0.08}\) & \checkmark &\(97.90\pm{0.10}\) & \checkmark\\
    MNIST & CNN & \(98.81\pm{0.14}\) & \checkmark &\(99.04\pm{0.07}\) & \checkmark\\
    FMNIST & MLP & \(85.64\pm{0.22}\) & \checkmark & \(86.30\pm{0.52}\) & \checkmark\\
    FMNIST & CNN & \(87.59\pm{0.47}\) & \checkmark &\(89.22\pm{0.18}\) & \checkmark\\
    CIFAR10 & MLP & \(37.08\pm{0.37}\) & \checkmark &\(39.48\pm{1.69}\) & \checkmark\\
    CIFAR10 & CNN & \(84.03\pm{0.69}\) & \checkmark &\(84.54\pm{0.28}\) & \checkmark\\
    Malimg\(^{*}\) & CNN & \(91.06\pm{4.56}\) & \checkmark &\(93.28\pm{4.30}\) & \checkmark\\
    Names & RNN & \(56.02\pm{0.70}\) & \checkmark &\(65.82\pm{0.90}\) & \checkmark\\
    FI2010\(^{\dagger}\) & RNN & \(0.1371\pm{0.0024}\) & \checkmark &\(0.1612\pm{0.0033}\)& \checkmark\\
    \end{tabular}}
    \end{center}
    \scriptsize{\({*}\)F1 score, \({\dagger}\)Cohen's kappa score are reported, since the datasets a highly unbalanced}
\end{table}

In Table~\ref{tab:res_trans}, we report the evaluation results of the proposed NNT when applied to traditionally trained DL models. More specifically, we optimize the network using the SGD optimizer and then we apply the proposed transformation to acquire its non-negative isomorphic network reporting if the performance matches exactly to the original one. In each case, after performing hyper parameter search, we evaluate the best configuration obtained (e.g. learning rate) evaluating them in 5 evaluation runs and report the average and variance of the evaluation accuracy (or F1 and \(\kappa\) scores on highly unbalanced datasets). The proposed transformation leads to the exact same accuracy irrespective of applied architectures, dataset, and/or photonic configuration.  The proposed transformation enables us to exploit known techniques of optimization without constraining the model's parameters during training, which is shown to lead to performance degradation~\cite{8051252, Chorowski2015a, 8489216}, and, in turn, transform to its non-negative isomorphic before being deployed on photonic hardware.

\begin{table}
    \centering
     \caption{Non-negative post training after transformation using MLPs}  
      \label{tab:half_fcnn}
      \resizebox{0.55\linewidth}{!}{
    \begin{tabular}{l|ll|l}
         \textbf{Dataset}  & \thead{Regular\\+ CSGD} & \thead{Proposed\\+ CSGD} &  \thead{Proposed\\+ NNSGD}  \\
         \midrule
         \multicolumn{4}{c}{Photonic Sigmoid}\\
         \midrule
         MNIST  & \(11.00\pm{0.00}\) & \(\bm{97.31\pm{0.90}}\) & \(\underline{\bm{97.40\pm{0.06}}}\) \\
         FMNIST & \(10.00\pm{0.00}\) & \(\bm{84.77\pm{0.21}}\) & \(\underline{\bm{85.01\pm{0.12}}}\) \\
         CIFAR10 & \(10.00\pm{0.00}\) & \(\bm{36.34\pm{0.37}}\) & \(\underline{\bm{36.81\pm{0.33}}}\)\\
         \midrule
         \multicolumn{4}{c}{Photonic Sinusoidal}\\
         \midrule
         MNIST  &  \(09.74\pm{0.00}\)  & \(\bm{97.85\pm{0.10}}\) & \(\underline{\bm{97.86\pm{0.07}}}\) \\
         FMNIST  & \(10.00\pm{0.00}\) & \(\bm{86.62\pm{0.09}}\) & \(\underline{\bm{86.95\pm{0.17}}}\)\\
         CIFAR10 & \(10.00\pm{0.00}\) & \(\bm{40.96\pm{0.30}}\) & \(\underline{\bm{41.19\pm{0.38}}}\)\\
    \end{tabular}}
\end{table}

\subsection{Non-negative post training}

We also evaluate the proposed transformation and non-negative optimization method when applied to regular pre-trained models. More specifically, we train models using SGD optimizer, and after a few epochs we transform the trained network and continue the training on its isomorphic model in a non-negative manner. We compare the proposed non-negative optimization method to a baseline non-negative training used in literature~\cite{8051252}, given by the update rule \(\theta_{t} = max\{0, \theta_{t-1} - \eta(\partial J/\partial \theta_{t-1})\} \in \mathbb{R^{+}}\), applied on the SGD optimizer, denoted as Clipping Stochastic Gradient Descent (CSGD). 

In Table~\ref{tab:half_fcnn}, we report the average accuracy of the evaluation phase and the variance over 5 evaluation runs using MLPs. The proposed framework, including both transformation and NNSGD optimization, is reported in the 4th column. As baselines, we evaluate CSGD optimization used directly after pre-training (column 2) and after applying the proposed transformation (column 3). As shown, directly applying clipping to trained parameters is catastrophic since it leads to a huge loss of information, while the network is unable to recover from it. When applying the proposed transformation, the transformed parameters preserve the information obtained from the training while the models are further optimized with CSGD. Applying the proposed NNSGD method, the evaluation performance is slightly improved compared to the CSGD. Compared to the regular training using the same epochs of training, the performance obtained is highly competitive, with some cases even exceeding them (such as in photonic sinusoidal configuration on FMINST and CIFAR10 datasets). Finally, we provide an experimental analysis on parameter variance in the Appendix, which shed light on the destructive effects of not applying the proposed transformation as well as the benefits of applying NNSGD.  

\begin{table}
     \caption{Non-negative post training after transformation using CNNs}  
      \label{tab:half}
      \begin{center}
      \resizebox{\linewidth}{!}{
    \begin{tabular}{l|llll|ll}
         \textbf{Method}  & MNIST & FMNIST & CIFAR10 & Malimg\(^{*}\) & Names & FI2010\(^{\dagger}\) \\
         \midrule
         \multicolumn{7}{c}{Photonic Sigmoid}\\
         \midrule
         Proposed (CSGD) & \(96.85\pm{0.96}\) & \(83.90\pm{0.46}\) & \(74.30\pm{0.22}\) & \(93.53\pm{2.76}\) & \(53.94\pm{1.44}\) & \(0.1050\pm{0.0070}\)\\
         Proposed (NNSGD) & \(\bm{97.60\pm{1.83}}\) & \(\bm{84.87\pm{0.33}}\) & \(\bm{77.39\pm{0.56}}\) & \(\bm{96.60\pm{0.78}}\) & \(\bm{58.37\pm{1.05}}\) & \(\bm{0.1220\pm{0.0022}}\)  \\
         \midrule
         \multicolumn{7}{c}{Photonic Sinusoidal}\\
         \midrule
         Proposed (CSGD) & \(98.58\pm{0.12}\) & \(87.59\pm{0.20}\) & \(77.30\pm{0.73}\) & \(92.18\pm{4.52}\) & \(66.64\pm{0.83}\) & \(0.1434\pm{0.0014}\) \\
         Proposed (NNSGD) & \(\bm{99.04\pm{0.07}}\) & \(\bm{87.63\pm{0.19}}\) & \(\bm{79.95\pm{0.21}}\) & \(\bm{93.53\pm{4.30}}\) & \(\bm{67.14\pm{0.54}}\) & \(\bm{0.1840\pm{0.0018}}\)  \\
    \end{tabular}}
    \end{center}
    \scriptsize{\({*}\)F1 score, \({\dagger}\)Cohen's kappa score are reported, since the datasets a highly unbalanced}
\end{table}

We extend the evaluation of the proposed method in both CNNs and RNNs with the results reported in Table~\ref{tab:half}. Both optimization methods achieve competitive results with the regular training, while the proposed non-negative optimizer slightly improves the performance of the non-negative isomorphic models compared to CSGD. Interestingly, in the Malimg dataset, the proposed non-negative optimizer improves the performance of the model even compared to the regular training (\(\sim 5\%\) in case of photonic sigmoid).

\begin{table*}

    \centering
     \caption{Non-negative training from scratch in MLPs}  
      \label{tab:full_fcnn}
    \resizebox{0.75\linewidth}{!}{
    \begin{tabular}{l|ll|ll}
         \textbf{Dataset}  & \thead{Exp. Initialization\\+ CSGD} & \thead{Proposed\\+ CSGD} & \thead{Exp. Initialization\\+ NNSGD} & \thead{Proposed\\+ NNSGD}  \\
         \midrule
         \multicolumn{5}{c}{Photonic Sigmoid}\\
         \midrule
         MNIST & \(11.35\pm{0.00}\) &\(\bm{92.52\pm{0.52}}\) & \(11.35\pm{0.00}\) & \(\underline{\bm{93.34\pm{1.11}}}\) \\
         FMNIST & \(10.00\pm{0.00}\) &\(\bm{82.31\pm{1.29}}\) & \(10.00\pm{0.00}\) & \(\underline{\bm{84.09\pm{0.56}}}\)\\
         CIFAR10 & \(10.00\pm{0.00}\) &\(\bm{37.25\pm{1.57}}\)  & \(10.00\pm{0.00}\) & \(\underline{\bm{37.61\pm{1.76}}}\)\\
         \midrule
         \multicolumn{5}{c}{Photonic Sinusoidal}\\
         \midrule
         MNIST  & \(86.36\pm{1.39}\) &\(\bm{92.51\pm{0.78}}\) & \(35.31\pm{6.11}\) & \(\underline{\bm{94.14\pm{0.90}}}\)\\
         FMNIST & \(68.95\pm{3.71}\) &\(\bm{82.91\pm{0.55}}\)   & \(9.97\pm{0.02}\) & \(\underline{\bm{83.57\pm{0.26}}}\)\\
         CIFAR10 &  \(10.00\pm{0.00}\) &\(\bm{34.96\pm{2.44}}\) & \(10.00\pm{0.00}\)  & \(\underline{\bm{35.41\pm{1.43}}}\)\\ 
    \end{tabular}}
\end{table*}

\subsection{Non-negative training from scratch}

In the final set of experiments, we evaluate the proposed framework in fully non-negative training. The proposed transformation is applied after randomly initializing the weights using the default PyTorch initialization~\cite{he2015delving}. In Table~\ref{tab:full_fcnn}, we use as baseline the non-negative exponential initialization, used in~\cite{8051252}. Thus, four different non-negative approaches are evaluated: a) applying non-negative exponential initialization combined with CSGD optimizer, b) applying the proposed transformation along with the CSGD optimizer, c) applying non-negative exponential initialization combined with proposed NNSGD optimizer, and d) applying the proposed non-negative framework. The non-negative exponential initialization (columns 2 and 4) leads to unstable performance in contrast to the proposed NNT (columns 3 and 5) which leads to more robust non-negative training. Compering the proposed non-negative optimization method with the CSGD, we conclude that the proposed non-negative optimization leads to slightly better performance.

\begin{table}
     \caption{Non-negative training from scratch in CNNs and RNNs}  
      \label{tab:full}
      \begin{center}
      \resizebox{\linewidth}{!}{
    \begin{tabular}{l|llll|ll}
         \textbf{Method}  & MNIST & FMNIST & CIFAR10 & Malimg\(^{*}\) & Names & FI2010\(^{\dagger}\) \\
         \midrule
         \multicolumn{7}{c}{Photonic Sigmoid}\\
         \midrule
         Proposed (CSGD) & \(97.69\pm{0.20}\) & \(83.81\pm{0.85}\) & \(71.61\pm{1.04}\) & \(91.47\pm{4.37}\) & \(54.00\pm{2.13}\) & \(0.1338\pm{0.0045}\) \\
         Proposed (NNSGD) & \(\bm{98.14\pm{0.23}}\) & \(\bm{83.87\pm{0.19}}\) & \(\bm{72.16\pm{0.54}}\) & \(\bm{91.98\pm{4.00}}\) & \(\bm{56.68\pm{1.11}}\) & \(\bm{0.1498\pm{0.0112}}\) \\
         \midrule
         \multicolumn{7}{c}{Photonic Sinusoidal}\\
         \midrule
         Proposed (CSGD) & \(98.54\pm{0.14}\) & \(87.77\pm{0.36}\) & \(69.87\pm{1.73}\) & \(96.53\pm{0.42}\) & \(64.17\pm{0.14}\) & \(0.1732\pm{0.0175}\)\\
         Proposed (NNSGD) & \(\bm{98.71\pm{0.05}}\) & \(\bm{87.86\pm{0.36}}\) & \(\bm{75.16\pm{1.83}}\) & \(\bm{97.26\pm{0.31}}\) & \(\bm{68.70\pm{0.78}}\) & \(\bm{0.1789\pm{0.0102}}\)\\
    \end{tabular}}
    \end{center}
    \scriptsize{\({*}\)F1 score, \({\dagger}\)Cohen's kappa score are reported, since the datasets a highly unbalanced}
\end{table}

Due to the poor performance when non-negative exponential initialization is applied, in Table~\ref{tab:full} we report the evaluation performance only in cases where the proposed transformation is applied. The results confirm that the proposed NNSGD can generalize to CNNs and RNNs architectures and improve the average evaluation accuracy compared to CSGD when a non-negative training from scratch is required. Similar to non-negative post-training the performance obtained is highly competitive to regular training with some cases even exceeding them (such as in Malimg dataset and in RNNs).

\section{Conclusions}
\label{sec:conclusions}
We introduce a transformation method that acquires the non-negative isomorphic of a regular trained or untrained neural network. Along with the proposed non-negative optimization method, one can either train from scratch a non-negative neural network or continue training when needed. The experimental results confirm that the acquired non-negative isomorphic results in exactly the same performance as the regular one, while when combined with the proposed non-negative optimization leads to competitive accuracies in contrast to regular training.

\section{Limitations and future works}
The proposed transformation cannot be directly applied in RNNs with gated mechanisms, such as GRUs and LSTMs, and further adjustments are needed. Also, the shifted activation function introduces a subtraction that can be integrated into the hardware implementation since it shifts the transfer function on the positive side of \(x\)-axis. As a result, the proposed non-negative optimization method ensures non-negativity to parameters only. In order to acquire also non-negative intermediate values the proposed transformation should be applied anew after non-negative post-training, which can slightly increase the computational complexity. Furthermore, it should be noted that non-negative ANNs leverage advantages on interpretability since they allow us to accomplish part-based learning due to the elimination of canceling neurons resulting in additive data representation~\cite{8051252} which is conceptually tied to human cognition~\cite{lee1999learning}. As has been shown, non-negativity constraints significantly improve the human interpretation of ANNs through the visual representation~\cite{8051252,  Chorowski2015a} of models, giving us further insights that can be used to further improve the performance of DL models~\cite{8489216}. The wide range of applications that the non-negative ANNs can be applied highlighting the significance of such a research direction.

\newpage



\bibliography{bibliography}
\bibliographystyle{plain}
\end{document}



\maketitle

\section{Photonic Neural Networks Deployment}
\label{sec:pnn_deployment}

The deployment of photonic neural networks (PNNs) requires the employment of optically enabled mechanisms and photonic building blocks for realizing the respective neural layer signals and parameters. Input signals are typically imprinted into the optical domain using optical modulators, while weighting functionality generally requires the use of variable optical attenuation schemes that are responsible for multiplying the power level of the optical signal. Several approaches have been so far utilized in on-chip weight implementations of photonic neural layers, including i) tunable optical filtering mechanisms, like thermo-optic Micro-Ring-Resonators~\cite{tait2017neuromorphic} or thermo-optic Mach-Zehnder interferometers~\cite{Mourgias-Alexandris:22, 9489835}, ii) tunable waveguide absorption techniques, relying mainly on the use of electro-absorptive Phase-Change-Material (PCM)-based waveguide structures~\cite{cheng2017chip, feldmann2019all} or electro-absorption modulators (EAMs)~\cite{9605987}, and iii) tunable optical gain approaches, as has been shown via the use of Semiconductor Optical Amplifier (SOA) modules onto a monolithic InP platform~\cite{10.1117/12.2587668, 9748238}. Summation is then performed in the optical domain via i) wavelength multiplexers or couplers, utilized when incoherent neuromorphic photonic architectures are employed where every input signal is carried by different wavelengths~\cite{feldmann2019all, 9557801, shastri2021photonics}, and ii) interferometric combiners and couplers, utilized in the case of coherent architectures where a single wavelength is used for the entire network~\cite{shen2017deep, 2023963973, 9605987, 9748238}. Finally, the non-linear activation can be offered either by i) optoelectronic schemes, where a photodiode has to be used for producing an electrical current that is proportional to the optical power level of the weighted sum~\cite{feldmann2019all, 2023963973, 9489835}, or ii) all-optical non-linear modules, where optical non-linear physical processes are utilized, responding upon the optical power of the weighted sum signal~\cite{feldmann2019all, 2023963973, 9489835}.

In this work, we are using two photonic activation functions that correspond to different photonic configurations used for providing the non-linear behavior required by Artificial Neural Networks (ANNs). The first is the photonic sigmoid, \(g(z): \mathbb{R} \rightarrow [0,1]\), introduced in~\cite{phsigmoid}:
\begin{equation} 
  g({z}) = A_2 + \frac{A_1 - A_2}{1 + e^{({z}-{z}_0)/d}},
  \label{eq:sig}
\end{equation}
where the parameters were set to \(A_1=0.060\), \(A_2=1.005\), \(z_0 = 0.145\) and \(d=0.033\) according to the experimental observations of a real hardware implementation.

For the second activation function, the layout proposed in~\cite{tait2017neuromorphic} is used, which is based on a Mach-Zehnder Modulator (MZM)~\cite{Pitris:18} to appropriately modulate an optical signal, along with a diode~\cite{sinusoidal}. The behavior of activation, \(g(z): \mathbb{R} \rightarrow [0,1]\),  is described by the following transfer function:
\begin{equation}
  g({z}) =
  \begin{cases}
    0, & \text{if } {z} < 0 \\
    \sin^2{\frac{\pi}{2}{z}} & \text{if } 0 \leq {z} \leq 1 \\
    1, & \text{if } {z} > 1 \\
  \end{cases},
  \label{eq:sin}
\end{equation}
denoted as photonic sinusoidal activation function.

The simplest PNN implementation would require amplitude modulators for its inputs, meaning that in this case both the input signal modulators and the photonic weighting stage control just the amplitude of the optical field, resulting in non-negative networks~\cite{9006831}. Introducing sign information in a PNN has to incorporate a new physical dimension that can be used to correlating its state with the sign. Coherent photonic architectures offer a rather simple way of representing signed optical signals by correlating the sign to the phase of the propagating optical fields~\cite{9605987, Mourgias-Alexandris:22, 9489835}. However, this requires additional phase modulation circuitry both at the input signal generation stage and weighting stage, also requiring more complex circuitry at the non-linear activation stage in order to account for the sign information of the weighted optical sum at the activation unit, like an optical biasing scheme or a coherent receiver~\cite{9606046, 9748238, 9605987, 9557801}. In incoherent photonic architectures, where optical signals can be a priori considered as non-negative power signals, the sign information has been so far incorporated by employing the spatial or wavelength dimension for representing positive and negative numbers. Spatial separation of positive and negative optical signals obviously requires more on-chip area, but also requires the use of balanced photodetection schemes at the activation stage in order to process separately the sum of all positive and the sum of all negative values and then calculate their different in the electrical domain~\cite{tait2017neuromorphic}. The use of the wavelength dimension for representing sign information relies on the simple concept of devoting different wavelengths for positive and negative numbers; however, this obviously requires a 100\% increase in the available wavelength resources and lasers compared to a non-negative incoherent architecture, while still necessitating the employment of balanced photodetection at the activation stage. Those different implementation requirements are presented in Table~\ref{tab:compare_neurons}.

\begin{table}[]
    \caption{Table depicting the different hardware requirements for a N:1 photonic perceptron deployed as a non-negative, incoherent and coherent layout. The use of photodiodes has been considered as part of the non-linear activation stage for converting the signal to electrical form.}
    \centering
    \resizebox{0.65\linewidth}{!}{
    \begin{tabular}{l|l|l|l}
    \toprule
        & \textbf{Incoherent}$^*$ & \textbf{Coherent} & \textbf{Non-negative} \\
        \midrule
         Lasers &  2N & 1 & N\\
         Input Amplitude Modulators & N & N & N\\
         Input Phase Modulators & - & N & N\\
         Weight Amplitude Modulators & N & N & -\\
         Weight Phase Modulators & - & N & - \\
         Additional Biasing Circuit & - & YES & - \\
         Photodiodes & 2 & 1 & 1 \\
         \bottomrule
    \end{tabular}}
    
    \scriptsize{\(^{*}\)with sign encoded onto wavelength dimension}
    \label{tab:compare_neurons}
\end{table}
\section{Algorithmic Description}

The proposed framework is outlined in the Algorithms~\ref{alg:nn_tranformation},~\ref{alg:nn_forward} and~\ref{alg:nnsgd}. First, the proposed transformation is presented in the Algorithm~\ref{alg:nn_tranformation}. It receives as input the model parameters denoted as \(\{\bm{\theta}^{(k)} \in \bm{\Theta} | \text{ where } k = 1 \ldots n\}\), where each \(\bm{\theta}^{(k)}\) includes weights \(\bm{w}^{(k)} \in \mathbb{R}^{M_k \times N_k}\) and biases \(\bm{b}^{(k)} \in \mathbb{R}^{N_k}\) of \(k\)-th layer. Moreover, it receives also the input shifting points \(\bm{\alpha} \in \mathbb{R}^{n}_+\), which depend on the activation function used and range of input features. The transformation algorithm outputs the non-negative parameters \(\bm{\Theta}'\) of the isomorphic model, the activation shifting points \(c^{(k)} \in \mathbb{R}^{n}_+\) for every layer and the input rotation mask \(\bm{m}\). The input rotation mask records signs of the original weights that are used during the feedforward pass to rotate the inputs where needed. The transformation method processes every \(k\)-th layer of the network (lines 2-9) by calculating the \(\tilde{b}_j^{(k)} \in \mathbb{R}_+\) value for each neuron of the \(k\)-th layer (line 4) using the input shifting point \(\alpha^{(k)} \in \mathbb{R}_+\), the activation shifting point \(c^{k}\) (line 6) while iteratively computing the input rotation mask (line 5). In turn, for each neuron of the \(k\)-th layer (lines 7-9), the new non-negative bias values \(b'^{(k)}_j \in \mathbb{R}_+\) (line 8) and weights \(\bm{w}'^{(k)}_i \in \mathbb{R}^{M_k}_+\) (line 9) are calculated. It is worth noting that the network could be trained or not. Taking as input the trained parameters of an original model, the transformation algorithm results in an isomorphic model with the same outputs as the original one, but with non-negative parameters and intermediate values. Subsequently, this can be especially useful for applications such as neuromorphic ANNs since the model can be trained in a traditional manner (without any non-negativity constraints) and then it can be transformed into a non-negative equivalent model to be used in inference without any performance degradation.
\begin{algorithm}
\caption{Non-Negative Transformation (NNT) \label{alg:nn_tranformation}}    
  \SetKwData{Left}{left}\SetKwData{This}{this}
  \SetKwData{Up}{up}\SetKwFunction{Union}{Union}
  \SetKwFunction{FindCompress}{FindCompress}
  \SetKwInOut{Input}{Input}
  \SetKwInOut{Output}{Output}
  
  \Input{
    \(\bm{\Theta} \in \{\bm{W}, \bm{B}\}\): model's parameters, \\
    \(\bm{\alpha} \in \mathbb{R}^{k}_{+}\): input shifting points, \\ 
  }
  \Output{
    \(\bm{\Theta}' \in \{\bm{W}', \bm{B}'\}\): non-negative parameters, \\ 
    \(\bm{c} \in \mathbb{R}^{k}_{+}\): activation shifting points,\\ \(\bm{m}\): input rotation mask
  }

  \Begin{
    \For{\(k = 1 \ldots n \)  }{
        \For{\(i = 1 \ldots N_{k}\)}{
            \(\tilde{b}_i^{(k)} = b_i^{(k)} - \alpha^{(k)}\sum_{\{j |w_{ij}^{(k)} < 0\}}|w_{ij}^{(k)}|\)\;
            \( \bm{m}^{(k)}_i = sign(\bm{w}^{(k)}_i)\)
        }
        \(c^{(k)} = max\{|\tilde{b}_1|, \ldots, |\tilde{b}_{N_k}|\}\) \;
        
        \For{\(i = 1 \ldots N_{k}\) }{
            \(b^{'(k)}_{i} = \tilde{b}_j^{(k)} + c^{(k)}\)\;
            \(\bm{w}^{'}_i{}^{(k)} = \left[|w_{i1}^{(k)}|, \ldots, |w_{iM_{k}}^{(k)}|\right]^{\top} \)\;
        }
        
    }
    \Return \(\bm{\Theta}', \bm{c}, \bm{m}\)\;
}
\end{algorithm}

The resulting non-negative model can be used for inference using the non-negative forward pass (NNF) as presented in Algorithm~\ref{alg:nn_forward}. The difference between regular and non-negative forward pass is that the inputs that originally associated with negative weights should be flipped according to the given input shifting point and input rotation mask (line 4). In turn, the algorithm performs activation shifting according to the activation shifting points provided by the Algorithm~\ref{alg:nn_tranformation}. This ensures that the output of the non-negative layer will remain the same as the original layer's output.
\begin{algorithm}
\caption{Non-Negative Forward (NNF) \label{alg:nn_forward}}
  \SetKwData{Left}{left}\SetKwData{This}{this}
  \SetKwData{Up}{up}\SetKwFunction{Union}{Union}
  \SetKwFunction{FindCompress}{FindCompress}
  \SetKwInOut{Input}{Input}
  \SetKwInOut{Output}{Output}
  
  \Input{\(\bm{\Theta}_{nn}\): non-negative parameter,\\
    \(\bm{x} \in \mathbb{R}^{M_0}_+\): input,\\
    \(\bm{c} \in \mathbb{R}^{k}_{+}\): activation shifting points,\\
    \(\bm{\alpha}\): input shifting points,\\
    \(\bm{m}\): input rotation mask,
  }
  \Output{\(x^{(n+1)}\): network's output}
  
  \Begin{

    \(\bm{y}^{(0)} = \bm{x}\) \;
    \For{\(k = 1 \ldots n \)  }{
        \For{\(i=1 \ldots N_k\)}{
            \(z_i = \sum_{j=1}^{M_k}w_{ij}^{(k)}\left( m_{ij}^{(k)}y_j^{(k-1)} + \frac{1 - m_{ij}^{(k)}}{2}\alpha^{(k)}\right) + b_i^{(k)}  \)\;
              \vspace{0.1cm}
            \(y^{(k)}_i = g_c^{(k)}(z^{(k)}_i - c^{(k)}) \) \;
        }
    }
    \Return \(x^{(n+1)}\)\;
  }
\end{algorithm}

Using the NNF, the network can be further trained in a non-negative manner, by applying the proposed Non-negative Stohastic Gradient Descent (NNSGD) optimizer ensuring that the trainable parameters remain non-negative. More specifically, the non-negative training process is implemented as the regular training, by propagating the loss from output to input layer and updating the trainable parameters accordingly for each mini-batch. The NNSGD optimizer is presented in Algorithm~\ref{alg:nnsgd}, receiving the set of parameters for each layer. First, it calculates the loss (line 2) and then the correction steps of the weights and biases (lines 3-4) are calculated using the associated learning rates (\(\eta_{in} \in \mathbb{R}\) and \(\eta_{out} \in (0, 1] \)). Finally, the weights and biases are updated (line 5) accordingly, with the algorithm implementing the proposed multiplicative update. The algorithms presents non-negative transformation and training on Multilayer Perceptrons (MLPs), but without loss of generality can be extended to also applied in Recurrent (RNNs) and Convolutional (CNNs) Neural Networks.  
  
  

\begin{algorithm}
\caption{Non-Negative Stochastic Gradient Descent (NNSGD) \label{alg:nnsgd}}    
  \SetKwData{Left}{left}\SetKwData{This}{this}
  \SetKwData{Up}{up}\SetKwFunction{Union}{Union}
  \SetKwFunction{FindCompress}{FindCompress}
  \SetKwInOut{Input}{Input}
  \SetKwInOut{Output}{Output}
  
  \Input{
    \(\bm{t}, \bm{y}\): target and prediction,\\
    \(\bm{w}, \bm{b}\): weight and bias,\\
    \(J\): loss function,\\
    \(\eta_{in}, \eta_{out}\): learning rates
  }
  \Output{\(\bm{w}, \bm{b}\): new weight and biases}

  \Begin{
  \(\mathcal{L} = J(\bm{t}, \bm{y})\)\;
  
  \(\Delta\bm{w} = |\bm{w}| \tanh{\left(-\eta_{in} \frac{\partial\mathcal{L}}{\partial\bm{w}}\right)}\eta_{out}\)\;
  \vspace{0.1cm}
  \(\Delta\bm{b} = |\bm{b}| \tanh{\left(-\eta_{in} \frac{\partial\mathcal{L}}{\partial\bm{b}} \right)} \eta_{out}\)\;
    \vspace{0.1cm}
  \(\bm{w} = \bm{w} + \Delta\bm{w}, \bm{b} = \bm{b} + \Delta\bm{b}\)\;
  
  \Return \(\bm{w}, \bm{b}\)\;
  }
\end{algorithm}

\section{Experimental Setup}

We evaluated the proposed method in a wide range of Deep Learning (DL) tasks with various architectures based on photonic sigmoid and sinusoidal activation functions, presented in~\ref{sec:pnn_deployment}. More specifically, the proposed framework is applied to MLPs, CNNs, and RNNs to demonstrate the wide range of applications that can be used. The applied models are kept small according to the current capacity and limitations of neuromorphic hardware. Even if some of the architectures employed are beyond the current capacity of existing photonic neuromorphic hardware, we consider it crucial to examine the scalability and potential use of the proposed framework in similar but quite bigger architectures. Thus, traditional benchmarks from various topics, such as image classification, time-series forecasting, and natural language processing, are used as experimental test bed.

More precisely, the following datasets are included:
\begin{itemize}
    \item MNIST~\cite{mnist}: consist of handwritten digits, including 60,000 samples at training set and 10,000 at evaluation set. The digits have been size-normalized, centered in a fixed-size, and be flattened for the fully-connected architectures to one dimension, leading to 784 features per sample.
    \item Fashion MNIST (FMNIST)~\cite{fashionmnist}: consists of \(28 \times 28\) grayscale fashion images of 10 categories, including 60.000 and 10.000 samples in the train and evaluation set, respectively.  
    \item CIFAR10~\cite{cifar10}: consists of 60000 32x32 colour images in 10 classes, with 6000 images per class. There are 50000 training images and 10000 test images.
    \item Malimg~\cite{malimg}: comprises 25 malware classes with every family has varying number of samples with the total number be more than 9.000. The Malimg dataset contains various size images generated from malware binary files. 
    \item FI2010~\cite{2543}:  a high frequency financial time series limit order book dataset that consists of more than 4.000.000 limit orders which come from 5 Finnish companies. The task of the forecast is to predict the movement of the future mid-price after the next 10 time steps which can go down, up or remain stationary. The data processing schema and the evaluation procedure are extensively described in~\cite{8713851}. For experiments, dataset splits 1 to 5 are used.
    \item Names~\cite{names_dataset}: contains a few thousand surnames from 18 languages of origin. The aim of the classification is to predict from which language a name is based on spelling. 
\end{itemize}

For each of these datasets we employed different architectures. First, we apply image classification using fully-connected architectures. More specifically, the models consist of two fully connected hidden layers with 200 and 100 neurons each for both MNIST and FMNIST, with a network input size of 784. Similarly to MNIST, images for CIFAR10 were converted to grayscale and classified from 2 hidden layer neural network with each layer consisting of 1024 and 512 neurons and the networks input equals to 1024. For all MLP experiments, the batch size is set to 256 with the models optimized for 10 epochs in the MNIST dataset and 30 epochs in the CIFAR dataset. Cross-entropy loss is applied during training for all datasets. 

MNIST and FMNIST datasets are also used to demonstrate the capabilities of the proposed method in CNN architectures. For both datasets, four convolutional layers followed by two fully connected layers are used. The first two convolutional layers are consist of 32 filters of size \(3\times3\), where in the second layer, stride 2 is used. The following two convolutional layers consist of 64 filters of size \(3\times3\), where stride 2 is used in the latter. Finally, the last two fully connected layers are consist of 128 and 10 neurons each, where the output feature map, as extracted from convolutional layers, is fed to them after been flattened. The optimization procedure, for all different cases, is based on the Stochastic Gradient Descent (SGD) optimizer using cross-entropy as loss function. Furthermore, CIFAR10 is also used in the experiment set, applying a CNN composed of 4 convolutional layers followed by 2 fully connected layers on the 3 channels digital images. More precisely, two convolutional layers with 32 and 64 filters of size \(3\times3\) followed by \(2\times2\) average pooling layers are applied. Then, another two convolutional layers are employed with 128 and 256 filters of size \(3\times3\) and a final \(2\times2\) average pooling layer. In turn, the output of the convolutional layer is flattened and fed to a fully connected layer with 512 neurons, before been forwarded to the final classification layer. Both models are optimized for 30 epochs using 256 batch size and cross-entropy loss. In Malimg dataset, models composed from 3 convolutional layers followed by 2 fully connected layers. The first two convolutional layers consist of 30 and 20 filters size of \(3\times3\) followed by a \(2\times2\) average pooling layer. The next convolutional layer consists of 5 filters size of \(3\times3\) followed by an average pooling layer that outputs a 3 channel feature map size of \(20 \times 20\). The extracted output is fed on a fully connected hidden layer, with 1000 neurons, and in turn to the classification layer. 32 batch size is used and models are trained for 30 epochs employing cross-entropy loss. 


Finally, the recurrent architecture is used for FI2010 consist of a recurrent photonic layer with 32 neurons. The output of the recurrent layer is fed to two fully connected layers consisting of 512 and 3 neurons, respectively. The length of the time series that is fed to the model is 10 (current and the past 9 timesteps) and the models are optimized for 20 epochs using cross entropy loss. For the Names dataset, we use a simple RNN architecture, with the classification layer taking as input the features and hidden state of the model. The outputs of the classification layer is fed to the softmax activation. The hidden recurrent layer takes as input the features and hidden state and, in turn, outputs the hidden state. The aforementioned recurrent architecture is optimized from 100 epochs using the negative log likelihood loss.  
\section{Experimental Analyses}

Two additional experiments are presented in this section. More specifically, in the first analysis, we demonstrate the distributions of trainable parameters and intermediate values of the non-negative isomorphic model in reference to the original one. Additionally, we demonstrate the benefits of applying the proposed transformation in a trained network, followed by non-negative optimization methods, by reporting the variance of model during training.     

\subsection{Non-negative Transformation}
\begin{figure}
    \centering
    \includegraphics[width=0.6\linewidth]{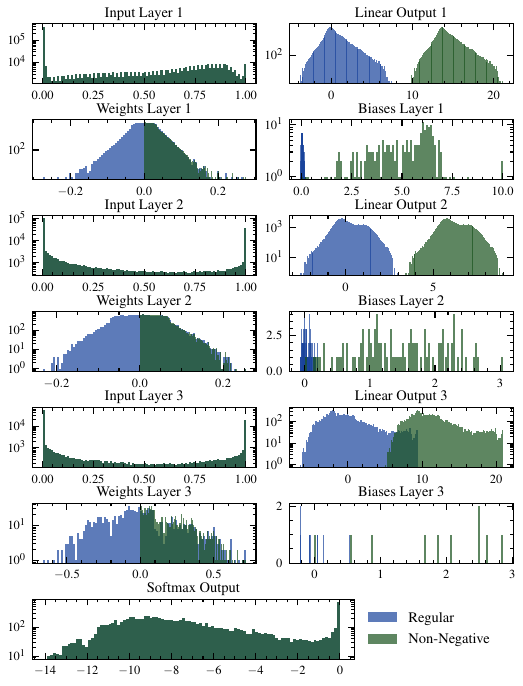}
    \caption{Histograms depict the distribution of trainable parameters, inputs and outputs of a non-negative transformed MLP in reference to the original network. Figure demonstrates a forward pass using both original and non-negative isomorphic models applying photonic sinusoidal activation in FMNIST dataset.}
    \label{fig:trans_hist}
\end{figure}

In Figure~\ref{fig:trans_hist}, the trainable parameter, input and output distributions of an MLP, which employs the photonic sinusoidal activation function and is trained on FMNIST dataset, are depicted in histograms. More specifically, it depicts both the original model, trained in a traditional manner, and its non-negative isomorphic equivelant after applying the proposed transformation. As shown, using the proposed transformation method, we are able to perform fully non-negative inference, since all the values involved during the inference remain positive. Furthermore, as depicted in the respective histograms, the outputs of every layer are the same for both regular and non-negative models. We should note that we also include the softmax output, even if it is not necessary during the inference to highlight the aforementioned property of the proposed transformation method.      

\subsection{Non-negative Post Training}
\begin{figure}
    \centering
    \includegraphics[width=\linewidth]{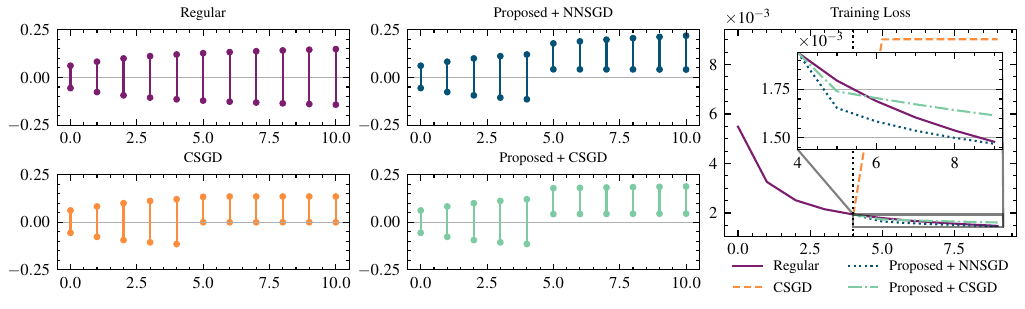}
    \caption{Depicts the variance of classification layer in photonic sigmoid architecture applied on FMNIST dataset. Additionally, in the left subfigure, the loss during the training is presented. Networks are pre-trained for 5 epochs with SGD and then they are either transformed with the proposed method or clipping is applied directly. The figure in the top left shows the variance of the regular training.}
    \label{fig:fmnist_variance}
\end{figure}

In Figure~\ref{fig:fmnist_variance}, the evolution of the variance, when non-negative post-training is applied, is presented in reference to regular training. More precisely, we report the variance of the classification layer of the photonic sigmoid architecture employed in the FMNIST dataset. In the upper left subfigure, the variance of the regular training is reported. The same network is transformed and trained in a non-negative manner applying the three evaluated methods. Thus after training the original model for 5 epochs, we a) apply Clipping Stochastic Gradient Descent (CSGD) optimizer directly, b) apply CSGD optimizer after employing the proposed transformation method, and c) apply the proposed NNSGD optimizer after employing the proposed transformation method, to model for another 5 epochs. As shown, applying directly clipping in a pre-trained model leads to variance reduction that results in loss of information obtained from the regular training and huge performance degradation. Applying the proposed transformation before clipping, it ensures that the variance of the weights will not diminish, allowing the training process to proceed smoothly and having enough distance from zero to increase the variance. However, it is observed that CSGD significantly delays the convergence of networks to local minima. This can be resolved with the proposed NNSGD optimizer that is able to converge faster to a better local minimum even compared to regular training.

\section{Hyperparameter Tuning}

For the hyperparameter tuning the ray framework~\cite{liaw2018tune} is used along with Nevergrad optimization platform~\cite{nevergrad}. More specifically, for each experiment, 30 different configurations are drawn from a given distribution using Nevergrad framework, such different learning rates for SGD (\(\eta\)) and CSGD (\(\eta_{nn}\)), while different inner (\(\eta_{in}\)) and outer learning (\(\eta_{out}\)) rates are sampled for NNSGD. In turn, the best configuration was selected according to the training metric, which is the training loss for all datasets. After that, the best set of parameters is used to apply for 5 evaluation runs, reporting the average value and variance of the evaluation metric in the corresponding paper's tables. 

\subsection{Trained model transformation}

In this first set of experiments, we conduct hyperparameter search in order to find the best learning rate, \(\eta\), for each case when regular training with SGD optimizer is employed. After applying 30 trail runs of the hyperparameter search algorithm for each different case, we used the best configuration of each case to report the average evaluation performance and variance in 5 evaluation runs in the tables presented in the paper. The values of learning rate during the hyperparameter searching is chosen by Nevergrad in each trial given a log-uniform distribution. To reduce search time, we introduce a grace epoch in which the Nevergrad algorithm can stop the trial if the current performance is significantly lower than in previous trials. The parameters of the hyperparameter search algorithm and the best configurations are presented for MLP architectures in Table~\ref{tab:regular_fcnn}. Similarly to MLP, we report the training parameters for CNNs and RNNs in Tables~\ref{tab:regular_cnn} and ~\ref{tab:regular_rnn}, respectively.

\begin{table}[]
    \centering
        \caption{Hyperparameter tuning configuration for MLPs used in regular training and their associated values as obtained after the hyperparameters search}
    \label{tab:regular_fcnn}
    \resizebox{0.8\linewidth}{!}{
    \begin{tabular}{l|l|l|l}
    \toprule
    & \textbf{MNIST} & \textbf{FMINST} & \textbf{CIFAR10}\\
    \midrule
    \textbf{Model size} &  178K & 178K & 1.3M \\
    
    \textbf{Epochs} & 10 & 10 & 30 \\
    \textbf{Batch Size} & 256 & 256 & 256 \\
    \textbf{Grace Epoch} & 7 & 7 & 20 \\
    \(\bm{\eta} \sim\) & LogUniform(\(1e-3, 1\)) & LogUniform(\(1e-3, 1\)) & LogUniform(\(1e-3, 1\)) \\
    \midrule      
    \multicolumn{4}{c}{\textbf{Photonic Sigmoid}}\\
    \midrule
    \textbf{SGD} & \(\eta=0.092\) & \(\eta=0.01\) & \(\eta=0.002\) \\
    \midrule      
    \multicolumn{4}{c}{\textbf{Photonic Sinusoidal}}\\
    \midrule
    \textbf{SGD} & \(\eta=0.5\) & \(\eta=0.16\) & \(\eta=0.047\) \\
    \bottomrule
    \end{tabular}}
\end{table}

\begin{table}[]
    \centering
        \caption{Hyperparameter tuning configuration for CNNs used in regular training and their associated values as obtained after the hyperparameters search}
    \label{tab:regular_cnn}
    \resizebox{\linewidth}{!}{
    \begin{tabular}{l|l|l|l|l}
    \toprule
    & \textbf{MNIST} & \textbf{FMINST} & \textbf{CIFAR10} & \textbf{Malimg}\\
    \midrule
    \textbf{Model size} &  197K & 197K & 3.6M  & 2M\\
    
    \textbf{Epochs} & 30 & 30 & 30 & 30 \\
    \textbf{Batch Size} & 256 & 256 & 256 & 32 \\
    \textbf{Grace Epoch} & 10 & 10 & 20 & 15 \\
    \(\bm{\eta} \sim\) & LogUniform(\(1e-3, 1\)) & LogUniform(\(1e-3, 1\)) & LogUniform(\(1e-3, 1\)) & LogUniform(\(1e-4, 1e-1\))\\
    \midrule      
    \multicolumn{5}{c}{\textbf{Photonic Sigmoid}}\\
    \midrule
    \textbf{SGD} & \(\eta=0.007 \) & \(\eta=0.01\) & \(\eta=0.026\) & \(\eta=0.0009\) \\
    \midrule      
    \multicolumn{5}{c}{\textbf{Photonic Sinusoidal}}\\
    \midrule
    \textbf{SGD} & \(\eta=0.5\) & \(\eta=0.056\) & \(\eta=0.06 \) & \(\eta=0.008\) \\
    \bottomrule
    \end{tabular}}
\end{table}

\begin{table}[]
    \centering
        \caption{Hyperparameter tuning configuration for RNNs used in regular training and their associated values as obtained after the hyperparameters search}
    \label{tab:regular_rnn}
    \resizebox{0.7\linewidth}{!}{
    \begin{tabular}{l|l|l}
    \toprule
    & \textbf{Names} & \textbf{FI2010}\\
    \midrule
    \textbf{Model size} &  34K & 24K \\
    
    \textbf{Epochs} & 100 & 20\\
    \textbf{Batch Size} & 128 & 128\\
    \textbf{Grace Epoch} & 40 & 10 \\
    \(\bm{\eta} \sim\)  & LogUniform(\(1e-4, 1e-1\)) & LogUniform(\(1e-4, 1e-1\)) \\
    \midrule      
    \multicolumn{3}{c}{\textbf{Photonic Sigmoid}}\\
    \midrule
    \textbf{SGD} & \(\eta=0.01\) & \(\eta=0.0013\) \\
    \midrule      
    \multicolumn{3}{c}{\textbf{Photonic Sinusoidal}}\\
    \midrule
    \textbf{SGD} & \(\eta=0.036\) & \(\eta=0.01\) \\
    \bottomrule
    \end{tabular}}
\end{table}

\subsection{Non-negative post training}

Similar hyperparameters search approach used also for the non-negative post training optimization. The models are optimized for \(T_r\) epochs using the SGD optimizer with the hyperparameters obtained from the above hyperparameter tuning process. The models are then transformed to their non-negative isomorphics, and in turn, non-negative training is performed for another \(T_{nn}\) epochs. For the MLP architectures, we evaluate 3 non-negative training approaches: a) applying CSGD after regular training, b) applying the proposed transformation and optimize the non-negative network using CSGD, and c) applying the proposed transformation and optimize the model with the proposed NNSGD algorithm. We performed a hyperparameter search for the hyperparameters associated with the different evaluated methods, using the Nevergrad framework. For each case, the hyperparameter search algorithm samples 30 different configurations from the given distributions before concluding to the best configuration, based on the training loss. Similarly to previous hyperparameter tuning process, we allow Nevergrad to stop a trial if the performance is not acceptable after the grace epoch. The best configuration is experimentally evaluated for 5 evaluation runs, and the results are reported in the paper. The searching distribution of each hyperparameter and best configuration setups are presented in Tables~\ref{tab:half_fcnn},~\ref{tab:half_cnn} and~\ref{tab:half_rnn}, associated with MLPs, CNNs and RNNs photonic architectures, respectively. 

\begin{table}[]
    \centering
        \caption{Hyperparameter tuning configuration for MLPs used in non-negative post training and their associated values as obtained after the hyperparameter searching}
    \label{tab:half_fcnn}
    \resizebox{0.9\linewidth}{!}{
    \begin{tabular}{l|l|l|l}
    \toprule
    & \textbf{MNIST} & \textbf{FMINST} & \textbf{CIFAR10}\\
    \midrule
    \(\bm{T_{r}}\) & 5 & 5 & 15 \\
    \(\bm{T_{nn}}\) & 5 & 5 & 15 \\
    \textbf{Grace Epoch} & 7 & 7 & 20 \\
    \(\bm{\eta_{nn}} \sim\) & LogUniform(\(1e-4, 1e-1\)) & LogUniform(\(1e-4, 1e-1\)) & LogUniform(\(1e-4, 1e-1\)) \\
    \(\bm{\eta_{in}} \sim\) & LogUniform(\(1e-3, 1\)) & LogUniform(\(1e-3, 1\)) & LogUniform(\(1e-3, 1\)) \\
    \(\bm{\eta_{out}} \sim\) & LogUniform(\(1e-3, 1\)) & LogUniform(\(1e-3, 1\)) & LogUniform(\(1e-3, 1\)) \\
    \midrule      
    \multicolumn{4}{c}{\textbf{Photonic Sigmoid}}\\
    \midrule
    \textbf{CSGD} & \(\eta_{nn}=0.0004\) & \(\eta_{nn}=0.0008\) & \(\eta_{nn}=0.0096\) \\
    \textbf{Proposed + CSGD} & \(\eta_{nn}=0.002\) & \(\eta_{nn}=0.0027\) & \(\eta_{nn}=0.0034\) \\
    \textbf{Proposed + NNSGD} & \(\eta_{in}=0.78, \eta_{out}=0.23\) & \(\eta_{in}=0.28, \eta_{out}=0.3\) & \(\eta_{in}=0.37, \eta_{out}=0.3\) \\
    \midrule      
    \multicolumn{4}{c}{\textbf{Photonic Sinusoidal}}\\
    \midrule
    \textbf{CSGD} & \(\eta_{nn}=0.0005\) & \(\eta_{nn}=0.0008\) & \(\eta_{nn}=0.0001\) \\
    \textbf{Proposed + CSGD} & \(\eta_{nn}=0.009\) & \(\eta_{nn}=0.008\) & \(\eta_{nn}=0.002\) \\
    \textbf{Proposed + NNSGD} & \(\eta_{in}=0.9, \eta_{out}=0.68\) & \(\eta_{in}=0.9, \eta_{out}=0.6\) & \(\eta_{in}=0.91, \eta_{out}=0.086\) \\
    \bottomrule
    \end{tabular}}
\end{table}

\begin{table}[]
    \centering
        \caption{Hyperparameter tuning configuration for CNNs used in non-negative post training and their associated values as obtained after the hyperparameters searching}
    \label{tab:half_cnn}
    \resizebox{\linewidth}{!}{
    \begin{tabular}{l|l|l|l|l}
    \toprule
    & \textbf{MNIST} & \textbf{FMINST} & \textbf{CIFAR10} & \textbf{Malimg}\\
    \midrule
    \(\bm{T_{r}}\) &  5 & 5 & 15  & 10\\
    
    \(\bm{T_{nn}}\) & 15 & 15 & 15 & 20 \\
    \textbf{Grace Epoch} & 10 & 10 & 20 & 15 \\
    \(\bm{\eta_{nn}} \sim\) & LogUniform(\(1e-4, 1e-1\)) & LogUniform(\(1e-4, 1e-1\)) & LogUniform(\(1e-5, 1e-2\)) & LogUniform(\(1e-4, 1e-1\))\\
    \(\bm{\eta_{in}} \sim\) & LogUniform(\(1e-3, 1\)) & LogUniform(\(1e-3, 1\)) & LogUniform(\(1e-3, 1\)) & LogUniform(\(1e-3, 1\))\\
    \(\bm{\eta_{out}} \sim\) & LogUniform(\(1e-3, 1\)) & LogUniform(\(1e-3, 1\)) & LogUniform(\(1e-3, 1\)) & LogUniform(\(1e-3, 1\))\\
    \midrule      
    \multicolumn{5}{c}{\textbf{Photonic Sigmoid}}\\
    \midrule
    \textbf{Proposed + CSGD} & \(\eta_{nn}=0.0024\) & \(\eta_{nn}=0.0035\) & \(\eta_{nn}=0.00002\) & \(\eta_{nn}=0.0008 \) \\
    \textbf{Proposed + NNSGD} & \(\eta_{in}=0.93, \eta_{out}=0.21\) & \(\eta_{in}=0.61\), \(\eta_{out}=0.082\) & \(\eta_{in}=0.25, \eta_{out}=0.19\) & \(\eta_{in}=0.4092, \eta_{out}=0.0787\) \\
    \midrule      
    \multicolumn{5}{c}{\textbf{Photonic Sinusoidal}}\\
    \midrule
    \textbf{Proposed + CSGD} & \(\eta_{nn}=0.04\) & \(\eta_{nn}=0.024\) & \(\eta_{nn}=0.00001\) & \(\eta_{nn}=0.0038\) \\
    \textbf{Proposed + NNSGD} & \(\eta_{in}=0.56, \eta_{out}=0.69\) & \(\eta_{in}=0.58, \eta_{out}=0.41\) & \(\eta_{in}=0.094, \eta_{out}=0.4\) & \(\eta_{in}=0.11, \eta_{out}=0.9633\) \\
    \bottomrule
    \end{tabular}}
\end{table}

\begin{table}[]
    \centering
        \caption{Hyperparameter tuning configuration for RNNs used in non-negative post training and their associated values as obtained after the hyperparameters searching}
    \label{tab:half_rnn}
    \resizebox{0.65\linewidth}{!}{
    \begin{tabular}{l|l|l}
    \toprule
    & \textbf{Names} & \textbf{FI2010}\\
    \midrule
    \(\bm{T_{r}}\)  & 50 & 10 \\
    
    \(\bm{T_{nn}}\) & 50 & 10\\
    \textbf{Grace Epoch} & 40 & 15 \\
    \(\bm{\eta_{nn}} \sim\) & LogUniform(\(1e-4, 1\)) & LogUniform(\(1e-4, 1\))\\   
    \(\bm{\eta_{in}} \sim\) & LogUniform(\(1e-4, 1\)) & LogUniform(\(1e-4, 1\)) \\
    \(\bm{\eta_{out}} \sim\) & LogUniform(\(1e-4, 1\)) & LogUniform(\(1e-4, 1\)) \\
    \midrule      
    \multicolumn{3}{c}{\textbf{Photonic Sigmoid}}\\
    \midrule
    \textbf{Proposed + CSGD}  & \(\eta_{nn}=0.34\) & \(\eta_{nn}=0.73\) \\
    \textbf{Proposed + NNSGD} & \(\eta_{in}=0.53, \eta_{out}=0.0001\) & \(\eta_{in}=0.084, \eta_{out}=0.0003\) \\
    \midrule      
    \multicolumn{3}{c}{\textbf{Photonic Sinusoidal}}\\
    \midrule
    \textbf{Proposed + CSGD} & \(\eta_{nn}=0.35\) & \(\eta_{nn}=0.53 \) \\
    \textbf{Proposed + NNSGD} & \(\eta_{in}=0.24, \eta_{out}=0.0001 \) & \(\eta_{in}=0.36, \eta_{out}=0.0023\) \\
    \bottomrule
    \end{tabular}}
\end{table}

\subsection{Non-negative training from scratch}

At the final set of experiments, we train from scratch non-negative neural networks. More precisely, for MLPs, we evaluate four different approaches: a) a non-negative exponential initialization combined with CSGD optimizer, b) the proposed transformation method combined with CSGD optimizer, c) a non-negative exponential initialization combined with the NNSGD optimizer and d) the proposed transformation method along with the NNSGD optimizer. For the baseline non-negative initialization (cases a and c), we used the non-negative initialization used in~\cite{Chorowski2015a}, in which authors drawn positive weights from an exponential distribution given by:
\begin{equation}
    T = \frac{\ln{(U(0, 1))}}{\lambda} \in \mathbb{R}^{+},
\end{equation}
where \(U(0, 1)\) is a uniform distribution between \((0, 1\)) and \(\lambda=100\). The proposed transformation is applied after the Kaiming initialization~\cite{he2015delving}. We perform a hyperparameter search to find the optimal learning rate (\(\eta_{nn}\)) for CSGD and the associated learning rates (\(\eta_{in}\) and \(\eta_{out}\)) for the proposed NNSGD optimizer. We conducted 30 experimental trials to find the best configuration setup based on training loss for every different case and architecture. The configuration of the hyperparameter searching are presented in Table~\ref{tab:full_nn_fcnn}, while the evaluation performance for the best configuration setup of every evaluated case is reported in the paper. Due to the poor performance of models when initialized with the exponential distribution, for the rest of the experiments, we evaluated only the non-negative optimization methods when combined with the proposed transformation. The parameters of the hyperparameter search algorithm are reported in Tables~\ref{tab:full_nn_cnn} and ~\ref{tab:full_nn_rnn} along with the best configuration for CNNs and RNNs, respectively.

\begin{table}[]
    \centering
        \caption{Hyperparameter tuning configuration for MLPs used in fully non-negative training and their associated values as obtained after the hyperparameters searching}
    \label{tab:full_nn_fcnn}
    \resizebox{0.9\linewidth}{!}{
    \begin{tabular}{l|l|l|l}
    \toprule
    & \textbf{MNIST} & \textbf{FMINST} & \textbf{CIFAR10}\\
    \midrule
    \textbf{Epochs} & 10 & 10 & 30 \\
    \textbf{Grace Epoch} & 7 & 7 & 20 \\
    \(\bm{\eta_{nn}} \sim\) & LogUniform(\(1e-4, 1e-1\)) & LogUniform(\(1e-4, 1e-1\)) & LogUniform(\(1e-4, 1e-1\)) \\
    \(\bm{\eta_{in}} \sim\) & LogUniform(\(1e-3, 1\)) & LogUniform(\(1e-3, 1\)) & LogUniform(\(1e-3, 1\)) \\
    \(\bm{\eta_{out}} \sim\) & LogUniform(\(1e-3, 1\)) & LogUniform(\(1e-3, 1\)) & LogUniform(\(1e-3, 1\)) \\
    \midrule      
    \multicolumn{4}{c}{\textbf{Photonic Sigmoid}}\\
    \midrule
    \textbf{Exp. Init. + CSGD} & \(\eta_{nn}=0.0001\) & \(\eta_{nn}=0.0002\) & \(\eta_{nn}=0.0001\) \\
    \textbf{Proposed + CSGD} & \(\eta_{nn}=0.0067\) & \(\eta_{nn}=0.0091\) & \(\eta_{nn}=0.0049\) \\
    \textbf{Exp. Init. + NNSGD} & \(\eta_{in}=0.17, \eta_{out}=0.41\) & \(\eta_{in}=0.91, \eta_{out}=0.06\) & \(\eta_{in}=0.075, \eta_{out}=0.151\) \\
    \textbf{Proposed + NNSGD} & \(\eta_{in}=0.64, \eta_{out}=0.6\) & \(\eta_{in}=0.28, \eta_{out}=0.91\) & \(\eta_{in}=0.5, \eta_{out}=0.5\) \\
    
    \midrule      
    \multicolumn{4}{c}{\textbf{Photonic Sinusoidal}}\\
    \midrule
    \textbf{Exp. Init. + CSGD} & \(\eta_{nn}=0.16\) & \(\eta_{nn}=0.3\) & \(\eta_{nn}=0.0001\) \\
    \textbf{Proposed + CSGD} & \(\eta_{nn}=0.034\) & \(\eta_{nn}=0.05\) & \(\eta_{nn}=0.024\) \\
    \textbf{Exp. Init. + NNSGD} & \(\eta_{in}=0.99, \eta_{out}=0.69\) & \(\eta_{in}=0.058, \eta_{out}=0.88\) & \(\eta_{in}=0.038, \eta_{out}=0.097\) \\
    \textbf{Proposed + NNSGD} & \(\eta_{in}=1, \eta_{out}=1\) & \(\eta_{in}=0.63, \eta_{out}=0.5\) & \(\eta_{in}=0.86, \eta_{out}=0.68\) \\
    \bottomrule
    \end{tabular}}
\end{table}

\begin{table}[]
    \centering
        \caption{Hyperparameter tuning configuration for CNNs used in fully non-negative training and their associated values as obtained after the hyperparameters searching}
    \label{tab:full_nn_cnn}
    \resizebox{\linewidth}{!}{
    \begin{tabular}{l|l|l|l|l}
    \toprule
    & \textbf{MNIST} & \textbf{FMINST} & \textbf{CIFAR10} & \textbf{Malimg}\\
    \midrule
    \textbf{Epochs} &  20 & 20 & 30  & 30\\
    \textbf{Grace Epoch} & 10 & 10 & 20 & 15 \\
    \(\bm{\eta_{nn}} \sim\) & LogUniform(\(1e-4, 1e-1\)) & LogUniform(\(1e-4, 1e-1\)) & LogUniform(\(1e-5, 1e-1\)) & LogUniform(\(1e-4, 1e-1\))\\
    \(\bm{\eta_{in}} \sim\) & LogUniform(\(1e-3, 1\)) & LogUniform(\(1e-3, 1\)) & LogUniform(\(1e-3, 1\)) & LogUniform(\(1e-3, 1\))\\
    \(\bm{\eta_{out}} \sim\) & LogUniform(\(1e-3, 1\)) & LogUniform(\(1e-3, 1\)) & LogUniform(\(1e-3, 1\)) & LogUniform(\(1e-3, 1\))\\
    \midrule      
    \multicolumn{5}{c}{\textbf{Photonic Sigmoid}}\\
    \midrule
    \textbf{Proposed + CSGD} & \(\eta_{nn}=0.004\) & \(\eta_{nn}=0.004\) & \(\eta_{nn}=0.002\) & \(\eta_{nn}=0.005\) \\
    \textbf{Proposed + NNSGD} & \(\eta_{in}=0.39, \eta_{out}=0.63\) & \(\eta_{in}=0.1\), \(\eta_{out}=0.11\) & \(\eta_{in}=0.28, \eta_{out}=0.38\) & \(\eta_{in}=0.26, \eta_{out}=0.69\) \\
    \midrule      
    \multicolumn{5}{c}{\textbf{Photonic Sinusoidal}}\\
    \midrule
    \textbf{Proposed + CSGD} & \(\eta_{nn}=0.039\) & \(\eta_{nn}=0.024\) & \(\eta_{nn}=0.0034\) & \(\eta_{nn}=0.01\) \\
    \textbf{Proposed + NNSGD} & \(\eta_{in}=0.83, \eta_{out}=0.78\) & \(\eta_{in}=0.77, \eta_{out}=0.56\) & \(\eta_{in}=0.71, \eta_{out}=0.37\) & \(\eta_{in}=0.34, \eta_{out}=0.37\) \\
    \bottomrule
    \end{tabular}}
\end{table}

\begin{table}[]
    \centering
        \caption{Hyperparameter tuning configuration for RNNs used in fully non-negative training and their associated values as obtained after the hyperparameters searching}
    \label{tab:full_nn_rnn}
    \resizebox{0.65\linewidth}{!}{
    \begin{tabular}{l|l|l}
    \toprule
    & \textbf{Names} & \textbf{FI2010}\\
    \midrule
    \textbf{Epochs}  & 100 & 20 \\
    \textbf{Grace Epoch} & 40 & 15 \\
    \(\bm{\eta_{nn}} \sim\) & LogUniform(\(1e-4, 1\)) & LogUniform(\(1e-4, 1\))\\   
    \(\bm{\eta_{in}} \sim\) & LogUniform(\(1e-4, 1\)) & LogUniform(\(1e-4, 1\)) \\
    \(\bm{\eta_{out}} \sim\) & LogUniform(\(1e-4, 1\)) & LogUniform(\(1e-4, 1\)) \\
    \midrule      
    \multicolumn{3}{c}{\textbf{Photonic Sigmoid}}\\
    \midrule
    \textbf{Proposed + CSGD}  & \(\eta_{nn}=0.0024\) & \(\eta_{nn}=0.0005\) \\
    \textbf{Proposed + NNSGD} & \(\eta_{in}=0.54, \eta_{out}=0.014\) & \(\eta_{in}=0.87, \eta_{out}=0.11\) \\
    \midrule      
    \multicolumn{3}{c}{\textbf{Photonic Sinusoidal}}\\
    \midrule
    \textbf{Proposed + CSGD} & \(\eta_{nn}=0.0086\) & \(\eta_{nn}=0.0033\) \\
    \textbf{Proposed + NNSGD} & \(\eta_{in}=0.75, \eta_{out}=0.36\) & \(\eta_{in}=0.68, \eta_{out}=0.2\) \\
    \bottomrule
    \end{tabular}}
\end{table}


\bibliography{bibliography}
\bibliographystyle{plain}